# Al-SUBSTITUTION EFFECTS ON PHYSICAL PROPERTIES OF THE COLOSSAL MAGNETORESISTANCE COMPOUNS $La_{0.67}Ca_{0.33}MnO_3$

I.G. Deac[a,*], L. Giurgiu[b], A. Darabont[a], R.V. Tetean[a], M. Miron[a], E. Burzo[a]

[a]*Faculty of Physics, "Babes-Bolyai" University, 3400 Cluj-Napoca, Romania*
[b]*Institute of Izotopic and Molecular Technology, 3400 Cluj-Napoca, Romania*

We present a detailed study of the polycrystalline perovskite manganites $La_{0.67}Ca_{0.33}Al_xMn_{1-x}O_3$ ($x$ = 0, 0.1, 0.15, 0.5) at low temperatures and high magnetic fields, including electrical resistance, magnetization, ac susceptibility. The static magnetic susceptibility was also measured up to 1000 K. All the samples show colossal magnetoresistance behavior and the Curie temperatures decrease with Al doping. The data suggest the presence of correlated magnetic clusters near by the ferromagnetic transition. This appears to be a consequence of the structural and magnetic disorder created by the random distribution of Al atoms.

## 1. Introduction

In the last few years there has been great interest in the relation between the structural, magnetic, and transport properties of $La_{1-x}Ca_xMnO_3$ and related rare earth manganate perovskites, following the observation of large magnetoresistance in these materials.[1,4] The occurrence of ferromagnetism had been attributed to the double-exchange (DE) interaction between $Mn^{3+}$ and $Mn^{4+}$ ions, and the richness of the phase diagrams[4] were naturally considered as a manifestation of the strong couplings among the spin, charge and Jahn-Teller (JT) lattice distortions. This coupling has recently been shown to commonly result in

---

* Corresponding author. Tel.: +4-026-440-5300; fax: +4-026-459-1906.
  *E-mail address*: ideac@phys.ubbcluj.ro.

electronic phase separation between different magnetoelectronic states at low temperatures.[5,6]

Despite the long history of work on these materials, systematic investigation studies of the properties upon impurity substitutions at the Mn site are necessary.

The structural, electrical and magnetic properties of CMR materials can be drastically changed either by the A-site substitutions[1] or by substitutions at the Mn site with *3d* elements.[2,3] This kind of latter substitution leads to changes in the type of magnetic interactions in these systems.

Since the behavior of the Mn ions are responsible for the electrical and magnetic properties of this compound it is interesting to change the distance between Mn ions in the compound as well as the $Mn^{3+}/Mn^{4+}$ ratio and to introduce disorder in the Mn lattice. This can be done by Al substitution at the Mn site. It has no magnetic moment (with no *d* electrons), its atomic radius is smaller than the Mn one, and it has the same valence with the host Mn (+3). Al doped in LCMO (La-Ca-Mn-O), in LSMO (La-Sr-Mn-O) and PCMO (Pr-Ca-Mn-O) leads to the decrease of the Curie temperature and some changes in the magnetoresitive properties.[7,8,3]

In the compounds $La_{1-x}Ca_xMnO_3$, the resistivity is lowest (at 80 K) for *x* = 0.33, composition corresponding to the best ferromagnetism, while high resistivities correspond to the antiferromagnetic compositions.[1] It is interesting to introduce Al in the compound with *x* = 0.33 to see how the transport, structural and magnetic properties change upon this doping. We found that, in spite of high Al content[7], the system had a CMR behavior with a I-M transition even in the absence of the magnetic field.

## 2. Experimental

Polycrystalline samples with nominal composition $La_{0.67}Ca_{0.33}Al_xMn_{1-x}O_3$ (*x* = 0, 0.1, 0.15, 0.5) were prepared by standard ceramic reaction. The compounds were sintered in air at $1400^0C$ for 24 h.

The phase and lattice parameters of the compounds were determined by X-ray powder diffraction using CuKα radiation with a Brucker Advance D8 AXS diffractometer.

The resistivity and magnetization of the samples were measured in Quantum Design PPMS cryostats and MPMS superconducting quantum interference device (SQUID) magnetometer respectively. A multipurpose Oxford Instruments MagLab System 2000 was used for ac susceptibility measurements. We used a Weiss-Forrer equipment to measure the dc susceptibility in the range from 300 to 1000 K.

### 3. Results and Discussion

The X-ray diffraction patterns of $La_{0.67}Ca_{0.33}Al_xMn_{1-x}O_3$ showed that the compounds are mainly clean single phase, within the limit of experimental errors. The lattice parameters were obtained by Rietveld analysis, using the FULLPROF program and they are indicated in Table 1 together with some other data. The space group *Pnma* was used for all the samples. The unit-cell volume, continuously decrease from $V = 231.99$ Å$^3$ for the sample with $x = 0$ to $V = 221.62$ Å$^3$ for $x = 0.5$, and this decrease is consistent with the lower size of $Al^{3+}$.

Figure 1 shows the results of the Al-substitution effect on the resistivity of LCMO in zero (open symbols) and in a 9 T magnetic field (solid symbols). All samples have a sharply suppressed resistivity in the high magnetic field for $T < T_c$, and this large is presumably associated with a field-induced ferromagnetic metallic state. The maximum of these curves indicates the occurrence of the metal-insulator transition that appears near by the onset of long-range magnetic order. In the low temperature region ($T < 40$ K) a shallow minimum appeared, that seems to be indicative of the intergrain conductivity and of the associated intergrain magnetoresitivity.[8] The magnetoresitance increases, at $T_c$, from 390 % for $x = 0$ up to 2741 % for $x = 0.15$, and this is in agreement with some other reports.[7] For $x = 0.5$, the magnetoresistance decreases to 290 %.

As shown in figure 1, the width of the M-I transition broadens with decreasing $x$ suggesting a distribution of grain sizes and the presence of the intrinsic inhomogeneities (metallic and insulating domains coexist, i.e. electronic phase separation, in a chemically homogenous grain) that exist inside the grains. It is interesting to remark that in spite of the highest value of the resistivity above $T_c$, for the sample with $x = 0.5$, curiously, the residual resisitvity of this sample is lower than that of the sample with $x = 0.1$.

Figure 2 describes the thermal dependence of magnetization of our samples in 0.05 T magnetic field. The samples were zero-field-cooled (ZFC), and the data were taken on warming. All samples show ferromagnetic behavior with $T_c$ decreasing with increasing Al content.

The $T_c$'s were estimated from the $M(T)$ curves in 0.05 T and they are displayed in table 1. In Fig. 3 is shown the thermal dependence of $H/M$ for the sample with $x = 0.5$ in the temperature range from 4 K to 300 K to estimate the paramagnetic Curie temperature $\Theta_c \sim 80$ K, in 0.05 T. For $x = 0$, the difference between the theoretical saturation magnetization and the measured one (taken at 5 K and 4 T) is very small. For higher Al content the magnetization is smaller and it does not saturate in fields up to 4 T. Figure 4 shows the thermal behavior of magnetization for all the samples in 4 T. For the sample with $x = 0.5$ saturation was not found even in 7 T. This behavior, as well as the change of $T_c$ can be understood taking account that Al doping will increase the distance between the Mn ions in a phase separation scenario in which the fraction of PM and/or AF phases seems to increase with Al doping. We do not exclude also the existence of some isolated Mn ions at the grain boundaries.

The results of the ac susceptibility measurements supplied $T_c$ values very closed to those obtained from magnetization measurements. A rather sharp maximum in $\chi'(T)$ was found, bellow $T_c$, only for the sample with $x = 0.5$ and this is shown in figure 5. The frequency dependence (in the range from 100 to 10 kHz) of the real part of the ac susceptibility, $\chi'(T)$, is very weak, within the noise range that is characteristic for a ferromagnetic behavior.

The $Mn^{3+}$ concentration, $f_{Mn}^{3+}$ was estimated from dc susceptibility measurements in the region of high temperature: 500 – 1000 K, to avoid the occurrence of ferromagnetic cluster near by $T_c$. At high temperatures $\chi_{dc}(T)$ follows in all cases a FM Curie–Weiss temperature dependence, $\chi_{dc}(T) = C/(T-\Theta)$. From the linear behavior of $\chi_{dc}^{-1}(T)$ we determined the Curie constants to estimate the $Mn^{3+}$ fraction. The value of $f_{Mn}^{3+}$ for each sample is also indicated in Table 1. As can be seen, Al doped at Mn site reduced the $Mn^{3+}$ concentration since it substitute mainly for $Mn^{3+}$, being isovalent ions.

For lower temperatures region the curve $\chi_{dc}^{-1}(T)$ has a positive curvature, suggesting that ferromagnetic cluster began to create. For this region, $T_c < T \leq 300$ K, the static magnetic susceptibility, $\chi_{dc}$ was estimated from the tail of $M(T)/H$ curves taken in a 500 Oe magnetic field. The Curie constants were also estimated from the linear behavior of $\chi_{dc}^{-1}(T)$ curves. The high value of the Curie constant just above $T_c$ ($C \approx 6$ emu K/mol) for samples with $x < 0.1$ means that the effective moment is larger than that expected for free manganese ions in $La_{0.7}Ca_{0.3}MnO_3$ ($C = 2.629$ emu K/mol). This can be explained by the presence of magnetic clusters. For the sample with $x = 0.5$ the Curie constant is $C = 1.43$ emu K/mol, and $\chi_{dc}^{-1}(T)$ is linear in the range 120-300 K. This suggests that Al doped in this sample can avoid the creation of large magnetic clusters starting from temperatures very closed to $T_c$. This behavior nicely correlates with the very sharp metal-insulator transition seen in the $\rho(T)$ curve for this sample, suggesting a reduced distribution of very small FM domains. It seems that low Al doping can create a large distribution of FM domains bellow $T_c$ and thick barrier between grains. At high Al content the small FM domains can fill better the sample creating many percolation paths, decreasing the residual resistivity.[10,11,12] The behavior of the most doped sample (i.e. lower resistivity) seems to be an effect of the presence of very small FM clusters for $T < T_c$. Also, for high Al content the Jahn-Teller distortion of $Mn^{3+}$ ions is weakened and this can make the structure more symmetric[9] with a sharp M-I transition and a lower residual resistivity. For higher temperatures $T > T_c$, in the absence of FM metallic domains Al acts as a potential scattering center and it cuts off the magnetic interaction between the Mn ions.[8] This makes the resistivity higher and the sample more paramagnetic (with smaller FM clusters) with increasing Al content, for this temperature region.

## 4. Conclusions

This study shows that the doping of manganese sites by aluminium in $La_{0.7}Ca_{0.3}MnO_3$ does not destroy the CMR effect. It decreases both the transition temperatures $T_c$'s and the $Mn^{3+}$ fraction in the system. Al doping increases the distances between magnetic ions and creates disorder in the compounds, enhancing the amount of PM and probably AF phases at low temperatures. For low Al content ($x \leq 0.15$) both the *MR*, and residual resistance $\rho_0$, increase with increasing *x*. When $x = 0.5$, *MR* and $\rho_0$ reduce their values. This behavior can be seen as percolation conduction in a phase separation scenario.

**References**

1. For reviews see J. M. D. Coey, M. Viret, and S. von Molnar, Adv. Phys.48 (1999) 167; and the references therein.
2. K. Ghosh, S. B. Ogale, R. Ramesh, R. L. Greene, T. Venkatesan, K. M. Gapchup, Ravi Bathe, and S. I. Patil, Phys. Rev. B 59 (1999) 553;
3. S. Herbert, A. Maignan, C. Martin, and B. Raveau, Solid State Commun., 121 (2002) 229.
4. P. Schiffer, A. P. Ramirez, W. Bao, and S-W. Cheong, Phys. Rev. Lett. 75 (1995) 3336;
5. A. Moreo, S. Yunoki, and E. Dagotto, Science 283 (1999) 2034;
6. I.G. Deac, J. F. Mitchell, and P. Schiffer, Phys. Rev. B 63 (2001) 172408;
7. J. Blasco, J. García, J.M. de Teresa, M.R. Ibarra, J. Perez, P.A. Algarabel, C. Marquina and C. Ritter, Phys. Rev. B 55 (1997) 8905.
8. Y. Sawaki, K. Takenada, A. Osuka, R. Shiozaki and S. Sugai, Phys. Rev. B 61 (2000) 11588;
9. C. Yaicle, B. Raveau, A. Maignan, M Herivieu, Solid State Commun. 132 (2004) 487
10. I.G. Deac, S.V. Diaz,. B. G. Kim, S.-W. Cheong and P. Schiffer Phys. Rev. B 64, (2002) 174426.
11. M. Ziese, Rep. Prog. Phys. 65 (2002) 143.
12. M. Uehara, S. Mori, C. H. Chen, and S.-W. Cheong, Nature (London) 399, (1999) 560.

Table caption:

Table 1. Quantitative data for the $La_{0.67}Ca_{0.33}Al_xMn_{1-x}O_3$

Figure captions:

Figure 1. Temperature dependence of the resistivity in zero magnetic field (open symbols) and in $H = 9$ T (solid symbols) of the entire series of samples.

Figure 2. Zero field cooled magnetization as a function of temperature measured in $H = 0.05$ T for the $La_{0.67}Ca_{0.33}Al_xMn_{1-x}O_3$ system

Figure 3. Zero field cooled magnetization as a function of temperature measured in $H = 4$ T for the $La_{0.67}Ca_{0.33}Al_xMn_{1-x}O_3$ system

Figure 4. Temperature dependence of the inverse susceptibility of $La_{0.67}Ca_{0.33}Al_{0.5}Mn_{0.5}O_3$. The solid line represents a Curie-Weiss fitting to the data.

Figure 5. Temperature and frequency dependence of the real part ($\chi'$) of the a.c. susceptibility of $La_{0.67}Ca_{0.33}Al_{0.5}Mn_{0.5}O_3$.

# Table 1. Deac et al.

| Sample LCMO | $a$ (Å) | $b$ (Å) | $c$ (Å) | V (Å$^3$) | MR (%) | $\rho_0$ (Ω cm) x 10$^{-5}$ | $T_{m-i}$ (K) | $T_c$ (K) | $f_{Mn^{3+}}$ (%) |
|---|---|---|---|---|---|---|---|---|---|
| $x=0$ | 5.461 | 7.756 | 5.477 | 231.99 | 390 | 19.4 | 181.1 | 208.4 | 57 |
| $x=0.1$ | 5.451 | 7.709 | 5.471 | 229.99 | 685 | 84.8 | 145.6 | 176.9 | 46.2 |
| $x=0.15$ | 5.455 | 7.650 | 5.422 | 228.08 | 2741 | 4695 | 96.9 | 125.5 | 30.3 |
| $x=0.5$ | 5.375 | 7.619 | 5.411 | 221.62 | 290 | 64.6 | 107.9 | 101 | 14.2 |

Fig.1, Deac et al.

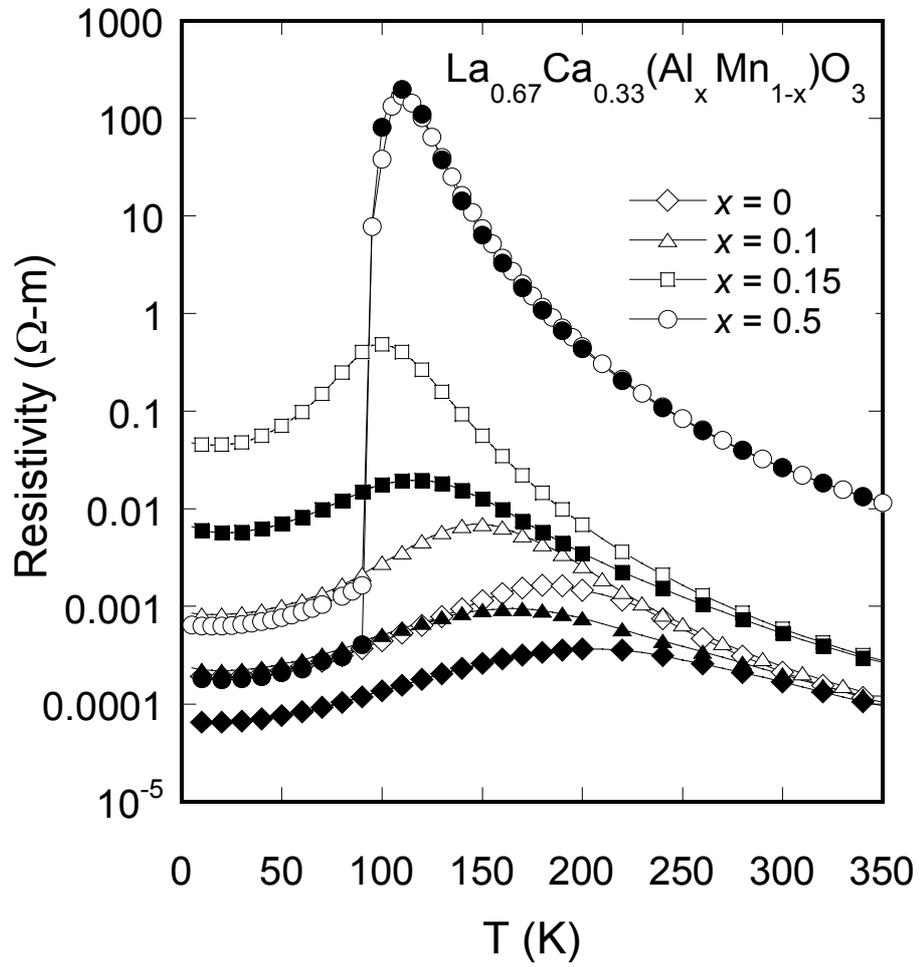

Fig.2, Deac et al.

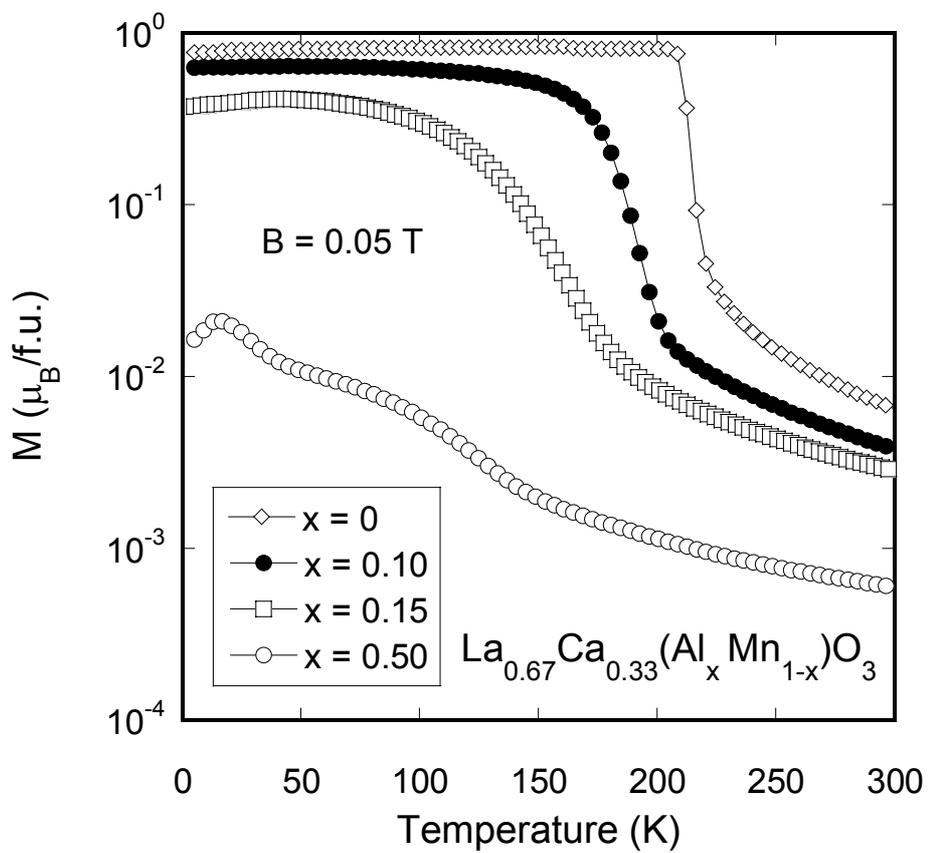

Fig.3, Deac et al.

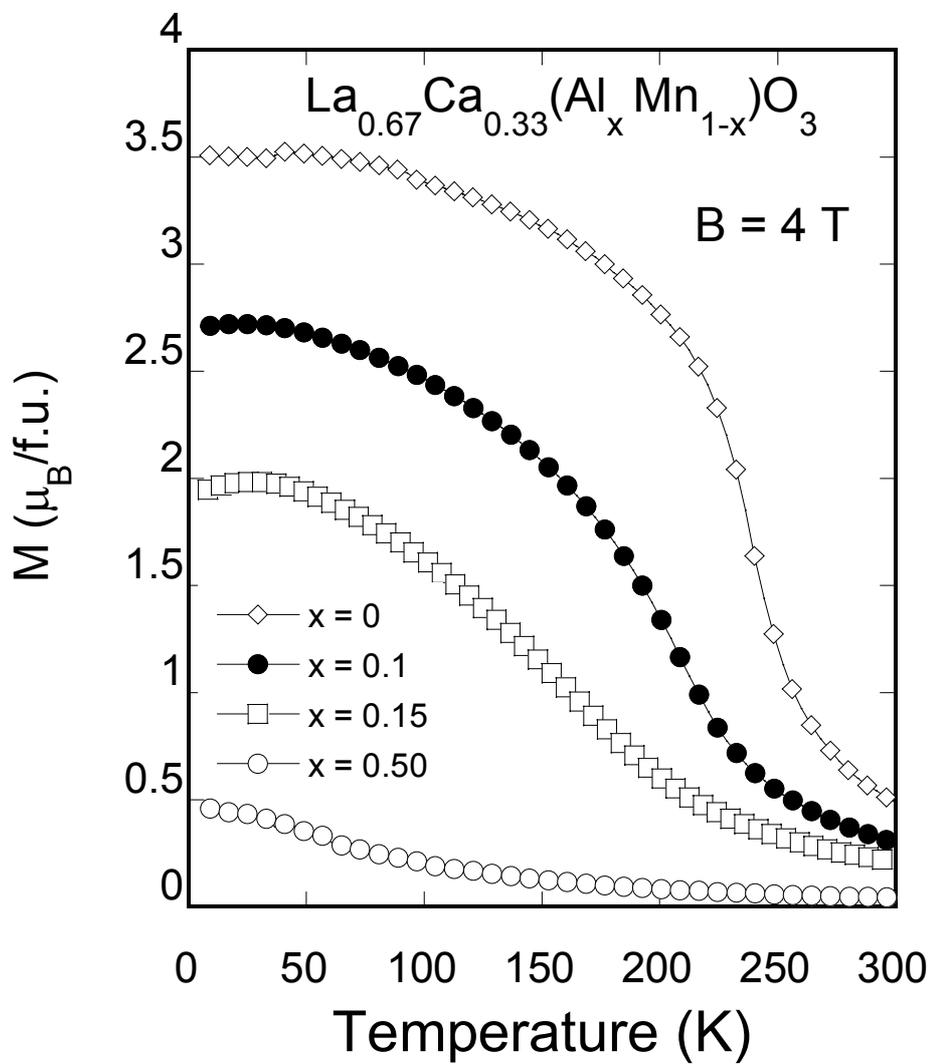

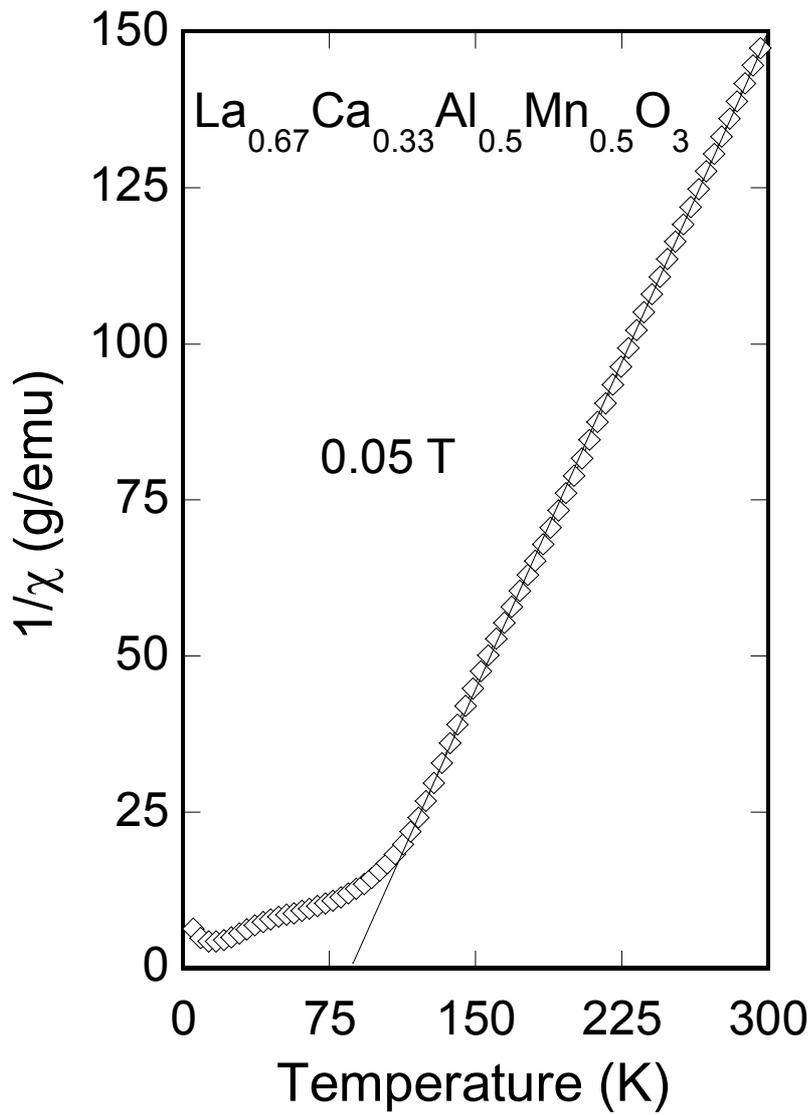

Fig.5, Deac et al.

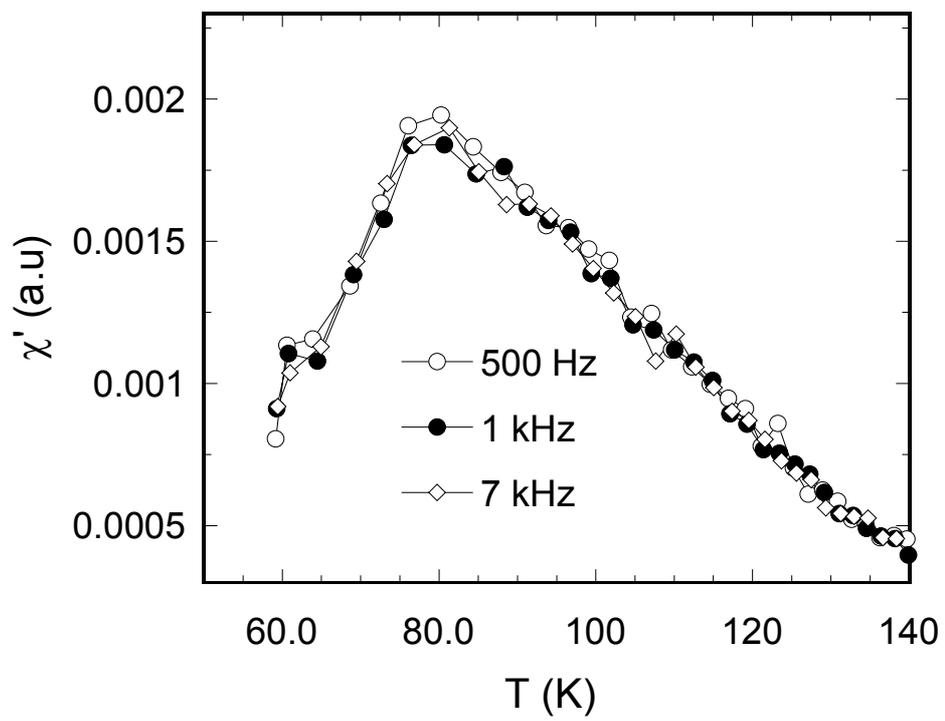